\documentclass[fleqn,twoside]{article}
\usepackage{espcrc2}
\usepackage{epsfig}

\newcommand{\AmS}{{\protect\the\textfont2
  A\kern-.1667em\lower.5ex\hbox{M}\kern-.125emS}}

\def\gs{\mathrel{
   \rlap{\raise 0.511ex \hbox{$>$}}{\lower 0.511ex \hbox{$\sim$}}}}
\def\ls{\mathrel{
   \rlap{\raise 0.511ex \hbox{$<$}}{\lower 0.511ex \hbox{$\sim$}}}}

\title{Deviations from tri-bimaximal mixing}

\author{W. Rodejohann\address{Max-Planck-Institut f\"ur Kernphysik, 
Postfach 103980, 69029 Heidelberg, Germany}\thanks{Work 
supported by the ERC under the Starting Grant 
MANITOP and the DFG in the Transregio 27.}}
       
\begin{document}

\begin{abstract}
Current data indicates that lepton mixing is very close to the 
tri-bimaximal mixing scenario. In general, however, one expects deviations 
from any mixing scenario. 
We discuss several examples to perturb 
tri-bimaximal mixing, namely soft breaking, renormalization 
group running and charged lepton corrections. 
We also present a convenient parametrization of the PMNS matrix 
which takes advantage of the apparently close to tri-bimaximal 
mixing pattern. 
Finally, it is discussed how to 
generate values of $U_{e3}$ which correspond to 
the recently found hint for non-zero $U_{e3}$.   
\vspace{-.5cm}

\vspace{1pc}
\end{abstract}

\maketitle

\section{Introduction}
Neutrino physics has entered the precision era. In particular, the entries 
of the unitary PMNS matrix $U$ are known with quite good precision. 
In general, 
\begin{equation} 
\label{eq:PDG}
U = R_{23} (\theta_{23}) \, \tilde R_{13}(\theta_{13}; \delta) \, 
R_{12}(\theta_{12}) \, P \, , 
\end{equation}
where $R_{ij}(\theta)$ is a rotation with angle $\theta$ around the $ij$-axis, 
e.g., 
\begin{equation}
R_{12}(\theta_{12}) =
\left(  
\begin{array}{ccc}
c_{12} & s_{12} & 0 \\
-s_{12} & c_{12} & 0 \\ 
0 & 0 & 1
\end{array}
\right) 
\end{equation}
and 
\begin{equation}
\tilde R_{13}(\theta_{13};  \delta) =
\left(  
\begin{array}{ccc}
c_{13} & 0 & s_{13} \, e^{-i \delta} \\
0 & 1 & 0 \\
-s_{13} \, e^{i \delta} & 0 & c_{13} \\ 
\end{array}
\right) .
\end{equation}
Here $c_{ij} = \cos \theta_{ij},\ s_{ij} = 
\sin \theta_{ij}$ with $\delta$ the CP-violating Dirac phase 
and $P = {\rm diag}(1, \, e^{-i\alpha_2/2}, \, e^{-i \alpha_3/2})$ 
contains the Majorana phases. 
Harrison, Perkins, and Scott first emphasized that the 
experimentally obtained mixing matrix is close to the simple tri-bimaximal 
mixing (TBM) form \cite{tri}:  
\begin{equation} \label{eq:UTBM}
\begin{array}{c}
U_{\rm TBM} = R_{23}(\pi/4) \, R_{12} (\theta_{\rm TBM}) 
\\
= 
\left( 
\begin{array}{ccc} 
\sqrt{\frac 23} & \sqrt{\frac 13} & 0 \\
-\sqrt{\frac 16} &  \sqrt{\frac 13} &  \sqrt{\frac 12} \\ 
\sqrt{\frac 16} &  -\sqrt{\frac 13} &  \sqrt{\frac 12}
\end{array}
\right)  ,
\end{array}
\end{equation}
where $\theta_{\rm TBM} = \sin^{-1} \sqrt{\frac 13}$. 
The apparent closeness to the TBM scenario invites to propose a parametrization 
of the PMNS matrix with TBM as the zeroth order expression. This 
``triminimal'' parametrization is \cite{PRW} (see also \cite{King}):  
\[ 
U =  R_{23}(\pi/4) \, R_{23}(\epsilon_{23}) \, 
\tilde R_{13}(\epsilon_{13}; \delta) \, 
R_{12}(\epsilon_{12}) \, R_{12} (\theta_{\rm TBM})\,, \nonumber 
\] 
with small $\epsilon_{ij}$. 
The virtues of triminimality are 
(i) each small $\epsilon_{ij}$ is responsible for 
the deviation of, and only of, $\theta_{ij}$ from its 
tri-bimaximal value;
(ii) the element $U_{e3}$ is $\sin \epsilon_{13} \, e^{-i\delta}$ and 
therefore possesses exactly the form as promulgated 
in the PDG-description of mixing matrices; 
(iii) is it obviously unitary.   

The other mixing observables besides $U_{e3}$ are 
\[ 
\begin{array}{c} \nonumber 
\sin^2 \theta_{12}  = \left( 
\cos \epsilon_{12} + \sqrt{2} \, \sin \epsilon_{12}
\right)^2 
\simeq \frac 13 + \frac{2\sqrt{2}}{3} \, \epsilon_{12} \,,\\ \nonumber 
\sin^2 \theta_{23} = \frac 12 + \sin \epsilon_{23} \, \cos \epsilon_{23} 
\simeq \frac 12 + \epsilon_{23}\,.
\end{array}
\]

\section{The mass matrix and deviations from TBM}
The Majorana mass matrix is defined as $m_\nu = U^\ast 
\, {\rm diag}(m_1, m_2, m_3) \, U^\dagger$ and takes a quite simple form 
in case of TBM: 
\[ 
(m_\nu)^{\rm TBM} = 
\left(
\begin{array}{ccc}
A & B & B \\
\cdot & \frac 12 (A + B + D) & \frac 12 (A + B - D) \\ 
\cdot & \cdot & \frac 12 (A + B + D)
\end{array}
\right) ,
\] 
where $A, B, D$ are functions of the masses and Majorana phases, see, e.g., 
\cite{AR}. 
\begin{figure}[ht]
\epsfig{file=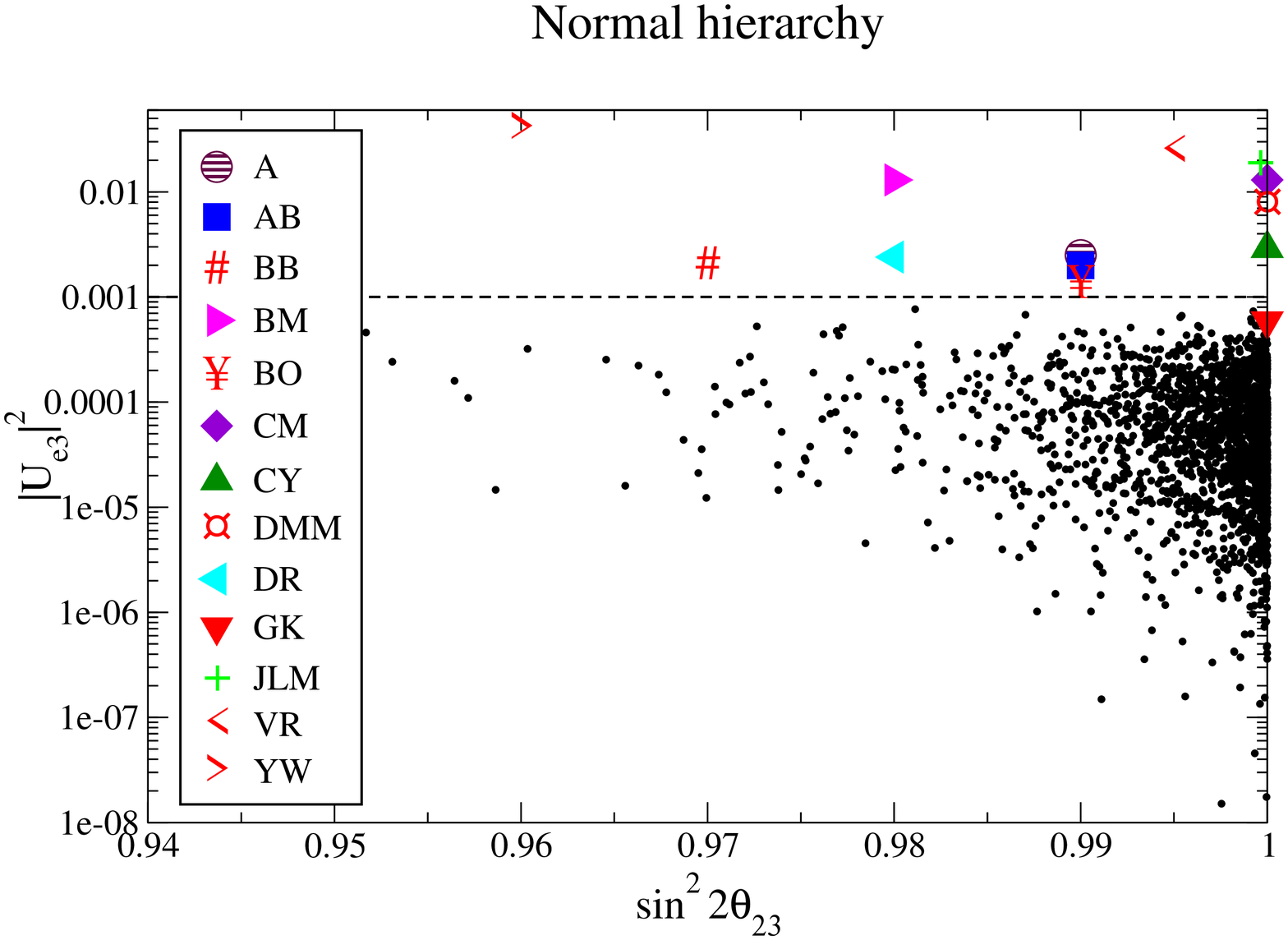,width=7.58cm,height=6cm}
\vspace{-.6cm}
\caption{Scatter plot of $|U_{e3}|^2$ against $\sin^2 2 \theta_{23}$ 
from a perturbed TBM mass matrix. Shown also are the results of successful 
$SO(10)$ GUTs. Taken from \protect\cite{AR}.}
\label{fig:AR}
\end{figure}
The given matrix is in fact obtained from a slightly modified 
(but physically identical) $U_{\rm TBM}$, 
which corresponds to choosing $\theta_{23} = - \pi/4$ in the 
standard PDG-description of eq.~(\ref{eq:PDG}). The most straightforward 
way to perturb TBM is to perturb every mass matrix entry, i.e., 
$(m_\nu)_{ee} = A \, (1 + \epsilon_{ee})$,  
$(m_\nu)_{e\mu} = B \, (1 + \epsilon_{e \mu})$ 
and so on. In total, there 
are six independent $\epsilon_{\alpha \beta}$, 
all of which are complex: 
$\epsilon_{\alpha \beta} = |\epsilon_{\alpha \beta}| 
\, e^{i \phi_{\alpha \beta}}$. We bound in the analysis, 
described in detail in \cite{AR}, $|\epsilon_{\alpha \beta}| 
\le 0.2$ and vary the 
phases between zero and $2\pi$. 
The implied deviations from TBM depend 
strongly on the neutrino mass ordering and scale. 
We find that the magnitude of $\theta_{13}$ is the most 
sensitive parameter. Atmospheric neutrino mixing is less sensitive, while 
$\theta_{12}$ has nothing to say at all. 
For a normal hierarchy of neutrino masses,  
$|U_{e3}|^2$ has a maximal value of $\frac 49 \, R \, |\epsilon|^2 \simeq 
7 \cdot 10^{-4}$, where $\epsilon$ is one of the 
$\epsilon_{\alpha \beta}$ and 
$R$ is the ratio of solar and atmospheric mass-squared differences. 
Fig.~\ref{fig:AR} shows a scatter plot resulting from the analysis. It 
compares broken TBM also with the predictions from 
13 successful $SO(10)$ GUTs, almost 
all of which predict $|U_{e3}|^2 \ge 10^{-3}$.  
In the inverted hierarchy, in turn, broken TBM can lead to 
$|U_{e3}|^2$ proportional to $|\epsilon|^2$ without suppression through 
$R$. It can therefore be as large as 0.01, while for 
quasi-degenerate neutrinos the 
complete allowed parameter space can be covered. 
All these features allow to 
distinguish GUTs from broken TBM if the mass ordering 
is known. 
The last two cases have 
an interesting correlation with neutrino-less double beta decay: if 
there is no cancellation in the effective mass 
$|(m_\nu)_{ee}|$,  
then $\theta_{13}$ is negligible, while for maximal 
cancellation $\theta_{13}$ can reach or exceed 
0.1. With ``no cancellation'' we denote the situation in which 
$|(m_\nu)_{ee}|/m_{\rm max} \simeq 1$, while ``maximal 
cancellations'' have occurred when 
$|(m_\nu)_{ee}|/m_{\rm max} \simeq\cos 2 \theta_{12}$, 
where $m_{\rm max}$ is the largest neutrino mass.

In principle included in the presented analysis are radiative corrections, 
because they can be estimated with good precision by 
zero $\epsilon_{ee, e\mu, \mu\mu}$ and  
$\epsilon_{e \tau} = \epsilon_{\mu \tau} = 
\frac 12 \, \epsilon_{\tau\tau} \equiv \epsilon_{\rm RG} 
= c \, \frac{m_\tau^2}{16 \pi \, v^2} \, \ln M_X/m_Z$ with 
$c = -\frac 32$ and $c = 1 + \tan^2 \beta$ in the SM and MSSM, 
respectively. In Ref.~\cite{DGR} 
it was shown that RG running leads 
to $\theta_{ij} = \theta_{ij}^{\rm TBM} + k_{ij}
 \, \epsilon_{\rm RG}$. Here $\theta_{ij}^{\rm TBM}$ are the mixing 
angles in case of TBM, and  
\[ 
\begin{array}{c} 
k_{12} = \frac{\sqrt{2}}{3}
\, \frac{\left| m_1 + m_2 \, e^{i \alpha_2}\right|^2}{m_2^2 - m_1^2} \, , \\
k_{23} = -
\left( 
\frac{2}{3} \, \frac{\left| m_2 + m_3 \, e^{i(\alpha_3 - \alpha_2)}
\right|^2}{m_3^2 - m_2^2}  
+ \frac{1}{3} \, \frac{\left| m_1 + m_3 \, e^{i\alpha_3}\right|^2}
{m_3^2 - m_1^2}  
\right)  ,\\
k_{13} = -\frac{\sqrt{2}}{3}
\left( 
\frac{\left| m_2 + m_3 \, e^{i(\alpha_3 - \alpha_2)}
\right|^2}{m_3^2 - m_2^2}  
- \frac{\left| m_1 + m_3 \, e^{i\alpha_3}\right|^2}
{m_3^2 - m_1^2}  
\right) .
\end{array} 
\] 
The sign and the size of the RG effects can be easily estimated now, depending 
of course on the neutrino mass ordering and hierarchy.  
To give one example, a value of $|U_{e3}| \simeq 0.1$ requires typically 
neutrino masses $\gs 0.1$ eV and $\tan \beta \gs 15$. 
In this case one requires 
a suppression of $k_{12}$ through $\alpha_2$ 
to keep $\theta_{12}$ under control. This 
is equivalent to maximal cancellation in the effective mass 
governing neutrino-less double beta decay.

\section{Deviating TBM with charged leptons}
Charged lepton corrections are quite popular when it comes to 
perturbing mixing scenarios \cite{AK,FPR,PR,HPR}. Using the 
relation $U = U_\ell^\dagger \, U_\nu$, it is usually 
assumed that $U_\nu$ corresponds 
to TBM and $U_\ell$ is, GUT-inspired, CKM-like.  
This means that it is given only by a 12-rotation with 
angle $\sin^{-1} \lambda$. In this case, one can show that 
\[ 
\begin{array}{c}
\sin^2 \theta_{12} 
\simeq \frac{ 1}{ 3} - \frac{ 2 }{ 3} 
\, {\lambda } \, {\cos \phi} \, ,
\\  
|U_{e3}| \simeq \frac{1}{\sqrt{2}} \,  
\lambda
~~,~~~J_{\rm CP} = \frac{\lambda }{ 6} 
\, {\sin \phi} \, ,
\\ 
\sin^2 \theta_{23} \simeq \frac{ 1}{ 2} 
- {\cal O}({ \lambda^2}) \, .
\end{array}
\]  
Here $\phi$ is a phase showing up in $U_\nu$ and 
$J_{\rm CP}$ is the Jarlskog invariant for neutrino oscillations.   
Hence, there is a direct correlation between 
$U_{e3}$, $\sin^2 \theta_{12}$ and CP violation. In particular, if 
$\lambda$ is large, e.g., the sine of the Cabibbo angle, then 
its contribution to $\sin^2 \theta_{12}$ is dangerously large unless we 
suppress it with $\phi = \pi/2$. In this case, the CP phase is 
maximal \cite{PR}. 

There is an alternative way of charged lepton corrections, namely when 
$U_\ell^\dagger$ corresponds to TBM and $U_\nu$ is CKM-like \cite{HPR}. 
In this case, there is a correlation in analogy to the one above, but now 
with atmospheric neutrino mixing, $|U_{e3}|$ and CP violation. It is 
harder to test because $|U_{e3}| \propto \lambda^2$ is smaller than in the 
previous case.

 \vspace{-.2cm}
\section{A hint for non-zero $\theta_{13}$}
Recently a hint for non-zero $U_{e3}$ 
emerged from global fits of neutrino data 
\cite{bari,others}. Ref.~\cite{bari} determines a $1\sigma$ range of 
$\sin^2 \theta_{13} = 0.016 \pm 0.010$, with a total significance of 
$1.6 \sigma$ for non-zero $\theta_{13}$.  
From the previous 
discussion\footnote{We should stress however that sizable values of 
$\theta_{13}$ are typical also for GUTs.}  
it is clear that values of $|U_{e3}| \simeq 0.1$ can be obtained by:  
\begin{itemize}
\item soft breaking of a TBM (or a $\mu$--$\tau$ symmetric)  
mass matrix in case of an inverted hierarchy 
or quasi-degenerate neutrinos. A characteristic signature in  
mass-related neutrino experiments, in particular neutrino-less 
double beta decay, 
follows in these regimes; 
\item charged lepton corrections to TBM from $U_\ell$ with $\lambda \simeq 
\sqrt{2} \, |U_{e3}|$,  
a number interestingly not too far away from the sine of the 
Cabibbo angle (or one third of it). In fact, the $1\sigma$ range 
given above corresponds to $\lambda = 0.18_{-0.07}^{+0.05}$. 
A characteristic correlation 
with solar neutrino mixing and CP violation 
($\delta \simeq \pi/2$ or $3\pi/2$) is obtained;
\item RG corrections to TBM in the MSSM with sizable 
$m_i \gs 0.1$ eV and $\tan \beta \gs 15$. Cancellations in the effective 
mass are required in order to suppress running of $\theta_{12}$. Note 
that MSSM-related lepton flavor violating processes typically 
grow with $\tan \beta$. 
\end{itemize}
It will be interesting to see if the currently weak hint survives or 
becomes more significant. In the likely case of one of the deviation 
possibilities being dominant, 
future precision data will then help us distinguishing them. 
In any case, the value of $|U_{e3}|$ used
here can be regarded as an interesting and testable 
benchmark scenario.\\

I wish to thank my co-authors for fruitful collaborations and 
the organizers of NOW 2008, E.~Fogli and 
G.L.~Lisi, for their invitation and for hosting 
a stimulating workshop.

\end{document}